# Querying hippocampal replay with subcortical inputs.

Adrien Peyrache [1]

[1] McGill University, Montreal Neurological Institute, Department of Neurology and Neurosurgery.

adrien.peyrache@mcgill.ca

## Abstract

During sleep, the hippocampus recapitulates neuronal patterns corresponding to behavioral trajectories during previous experiences. This hippocampal replay supports the formation of long-term memories. Yet, whether replay originates within the hippocampal circuitry or is initiated by extrahippocampal inputs is unknown. Here, I review recent findings regarding the organization of neuronal activity upstream to the hippocampus, in the head-direction (HD) and grid cell networks. I argue that hippocampal activity is under the influence of primary spatial signals, which originate from subcortical structures and set the stage for memory replay. In turn, hippocampal replay resets the HD network activity to select a new direction for the next replay event. This reciprocal interaction between the HD network and the hippocampus may be essential in providing meaning to hippocampal activity, specifically by training decoders of hippocampal sequences. Neuronal dynamics in thalamo-hippocampal loops may thus be instrumental for memory processes during sleep.

## Introduction

The role of the hippocampus and its associated structures in the formation of long-term explicit memories (of facts and events) has long been established [1]. Interestingly, the same circuits that are involved in memory are involved in representing the present location and the ability to navigate [2]. The hippocampus is the seat of a spatial cognitive map, as evidenced by the presence of "place cells", which fire when the animal is in a specific location in the environment [3] and allow it to navigate [4]. What is the relationship between navigation and long-term memory formation? In a seminal paper, David Marr wrote [5]:

"*The neocortex may be regarded as a structure which classifies the information presented to it. […] At a given moment, there will exist in an animal's brain information whose expression is now as sophisticated as the animal either requires or can provide.* **Further classification of the information may be carried out later** *but, at that moment, the animal needs simply to be able to store it in its present form*. [..] *the* **Regio Entorhinalis** *and the* **Regio Presubicularis** *prepare information from many different sources for its simple representation in the [hippocampus]*"

Prominent theories from Marr and followers posit a two-way dialogue between the hippocampus and the neocortex: encoding of neocortical states in the hippocampus during experience and spontaneous reactivation of hippocampal patterns during sleep [5–7], demonstrated by the observation that patterns of co-active place cells during navigation are replayed during subsequent sleep episodes [8,9]. Repetition of this activity would allow the neocortex to eventually store memories, even for episodes that have, by definition, happened only once. However, the mechanisms that determine the content of hippocampal replay remain unknown. Is the content governed entirely by intrahippocampal circuits and just decoded by the same regions that encode information during wakefulness, namely the entorhinal cortex and the presubiculum, to use more modern terminology? Or is the hippocampus still influenced by its inputs while in this "disengaged" state? Here, I review the sequence of physiological events that support replay and present hypothesis that subcortical signals, in particular those from the HD network, constrain hippocampal ensembles during sleep into activity patterns corresponding to specific travel directions.

## Thalamocortical dynamics during NREM sleep and their relationship with learning and memory

Since the seminal discovery of memory replay during non-rapid eye movement (NREM) sleep[8,9], efforts have been deployed to identify its neuronal and circuit basis. In the hippocampus, network activity during NREM is dominated by sharp wave-ripples (SWRs), high frequency (120-200 Hz) transient oscillations lasting about 100 ms [10]. SWRs are associated with the activation of neuronal sequences similar to those seen during previous experiences [9,10], which are instrumental in memory formation [11–13].

Beyond the hippocampus, NREM sleep is characterized by a dramatic shift in neuronal dynamics throughout the forebrain. Around the same time as the discovery of hippocampal replay, it was shown that neurons in thalamocortical networks fluctuate between activated UP and inactivated DOWN states, forming at the population level the so-called slow oscillation in the local field potentials (LFPs) [14]. Replay events are not independent this cortical activity. The occurrence of SWRs is influenced by the phase of the slow oscillation [15–17], with a marked decrease in SWR occurrence during thalamocortical DOWN states [17,18]. In turn, it was recently suggested that SWRs influence thalamocortical dynamics by resetting the phase of the slow oscillation [19].

DOWN-UP state fluctuations tend to be synchronous over large cortical networks, although localized fluctuations are often observed. The anterior thalamus, in particular the anterodorsal nucleus (ADN) of the thalamus, plays an essential role in synchronizing UP states in the posterior and medial cortex [20]. So far, the influence of thalamocortical inputs on hippocampal activity, especially on SWRs during NREM, have been mainly investigated in terms of overall population firing and LFP. It is only recently that the information conveyed by spontaneous activity upstream of the hippocampus has been decoded during NREM sleep.

## Attractor dynamics upstream of the hippocampus during wake and sleep

In order to make use of a map, one needs a compass to self-orient. In the brain, this is in the form of head direction (HD) neurons. HD neurons each fire for a specific direction of the head in the horizontal plane [21] (Fig. 1A). The HD signal can be considered the most fundamental spatial signal in the brain. While other spatial signals are in egocentric coordinates and must be inferred from e.g. visual inputs, the HD signal originates from subcortical circuitry which conveys information in allocentric coordinates. The HD signal originates in the brainstem and reaches the cortex via the ADN. While the functional neuroanatomy of the HD circuit has been the subject of a large body of experimental work (see refs [22] for review), this review will focus on the thalamocortical component of the HD system, linking ADN with its main cortical target - the post-subiculum (PoSub, or dorsal presubiculum).

The crucial property of the HD system is to represent a single direction at any given time. To ensure reliable signal transmission independent of the availability of particular sensory modalities (e.g. in light versus in darkness), a direction is encoded by a unique subset of similarly tuned HD neurons (Fig. 1B). From a computational perspective, the HD system is a canonical example of an attractor network [23] whereby internal circuits constrain neuronal population activity into a low dimensional subspace of possible states. For the HD system, these states are continuously distributed along a functional ring (Fig. 1C), which can be identified in by projecting multi-neuronal data onto a low dimensional map using dimensionality reduction and machine learning techniques [18,24] (Fig. 1C).

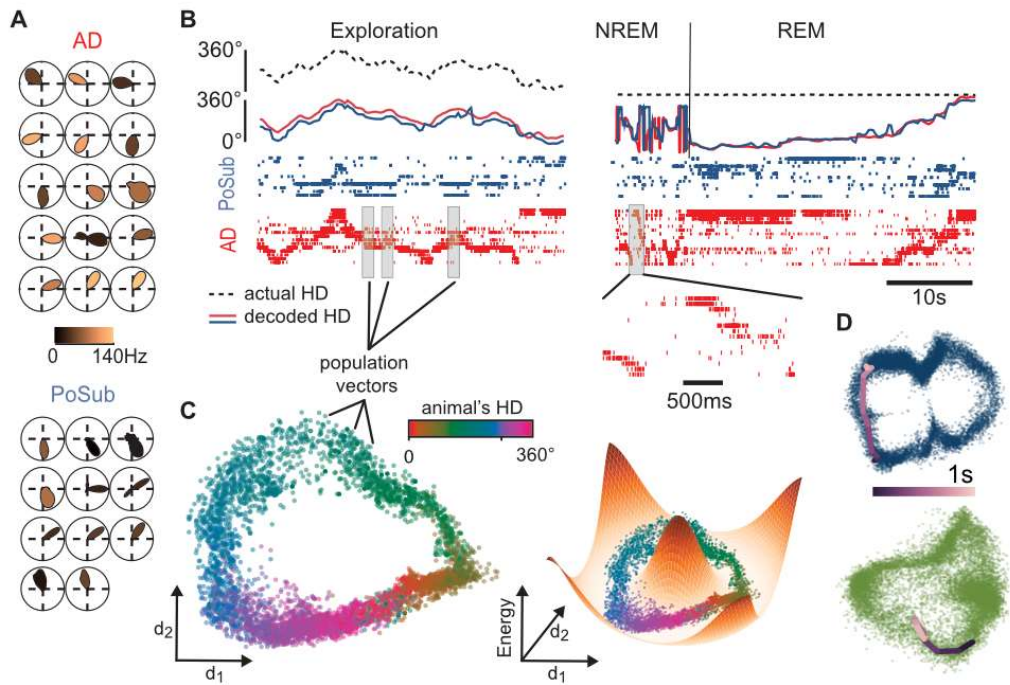

**Figure 1. Attractor dynamics in the HD network.**
A) Tuning curves (in polar plots) of simultaneously recorded HD Cells in the ADN-PoSub network. Color displays peak firing rates.
B) HD cell population activity across brain states (exploration, NREM and REM, from left to right). *Top*, actual animal's HD (dotted line) and Bayesian reconstruction from ADN and PoSub HD cell population in red and blue, respectively. Bottom, raster plots of HD cell activity, each row showing the spiking activity of an individual cell. HD cells are sorted by preferred direction separately for each structure.
C) *Left*, 2-dimensional projection of population activity during wakefulness. Each dot is the projection of a population vector, colored by the instantaneous animal's HD. Population vectors for similar animal's HD are close in projection space. *Right*, the organization of population activity along a functional ring may correspond to states of low energy in the system, as predicted from a ring attractor network.
D) Two-dimensional projection of population activity during wakefulness (top, blue) and REM (bottom, green). Example trajectories of HD cell population activity lasting one second during these two brain states are overlaid.
A,B adapted from ref [25]. C,D adapted from refs [18,24]

A key feature of an attractor network is that population activity maintains its internal organization irrespective of brain state. This should be particularly the case during REM sleep, when the patterns of electrical activity and the levels of neuromodulation are similar to wakefulness [26]. In fact, HD cell population activity during REM sleep is, at first sight, virtually indistinguishable from wakefulness, despite drifting around the ring in an apparently random fashion (Fig. 1D).

The medial entorhinal cortex (MEC) sits at a unique position between the head direction system and the hippocampus. It is one of the main output structures of the PoSub, and a majority of MEC neurons are modulated by the animal's HD [27]. It is also the main input and output structure of the hippocampus in the cortex, and grid cells in the MEC show regular firing in space such that their firing fields tessellate the environment with an hexagonal pattern [28]. The HD signal is certainly an essential input for the emergence of grid cells [29]. A large class of computational models of grid cell generation suggest they are endowed with the same kind of attractor dynamics as HD cells [30–32]. As predicted, the organization of grid cell population activity is also preserved during REM sleep [33–35]. These observations provide evidence for attractor dynamics within the grid cell network.

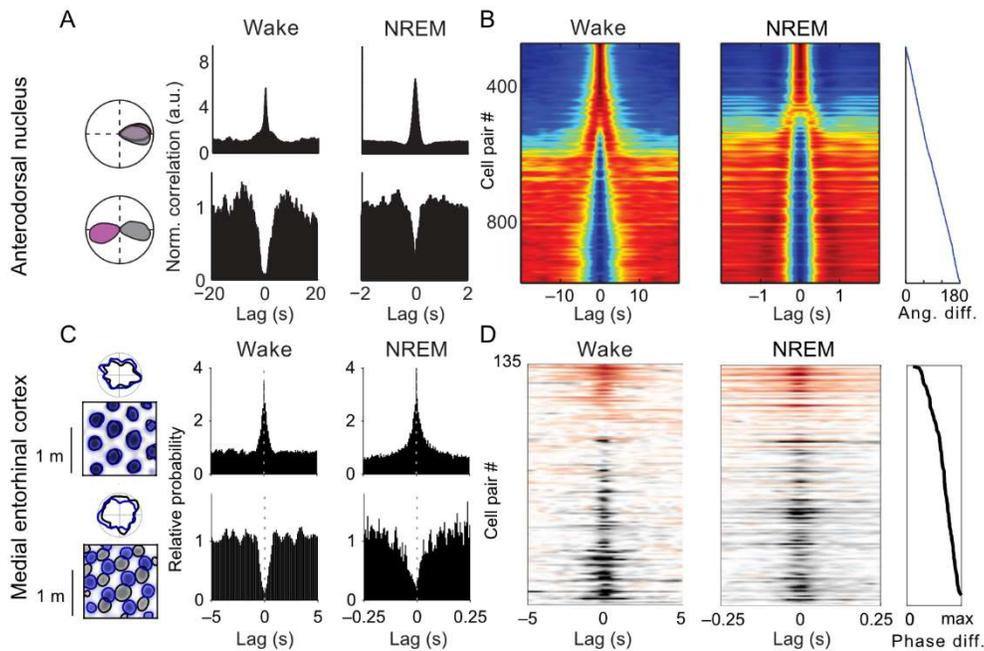

*Figure 2.* **Preserved coordinnation at fast temporal scale in the HD-grid network during NREM.**
A) Left, tuning curves of two HD cell pairs firing for the same (top) and opposite (bottom) directions. Right, spike train cross-correlograms of the same pairs during wakefulness and NREM. Note the change in time axis.
B) Cross-correlograms of multiple HD cell pairs (same as A but color-coded from blue to red showing minimal to maximal correlation values for each pair, respectively). Pairs are sorted by angular offset of preferred direction for each pair (right).
C-D) Same as A-B for pairs of simultaneously recorded grid cells of similar grid spacing (note that spike train coordination are computed with generalized linear models, not directly cross-correlograms). Grid cells may be in phase (C, top) and anti-phase (C, bottom). Cell pairs in D are sorted by phase difference and coordination is color coded from dark to red, showing minimal to maximal co-activation probability, respectively.
A-B adapted from ref [25]; C-D adapted with permission from ref [33].

Attractor dynamics in the HD-grid cell network, as evidenced by the high similarity between activity patterns during wakefulness and REM sleep, raises the possibility that neuronal coordination must be preserved during NREM sleep as well. At the population level, HD cell activity seems coordinated but drifts at a much faster pace than during wake/REM (Fig. 1B). Spike train cross-correlograms of HD cell pairs show that while the magnitude of zero-lag correlation between wake and NREM is preserved, they appear temporally compressed during NREM (Fig. 2A). Hence, the timescale at which the system remains in each position is an order of magnitude shorter than during wakefulness and REM sleep. The same observations were made for grid cells [33,35], which preserve their co-activation patterns while drifting at fast speed during NREM (Fig. 2B). This temporal compression of HD and grid neuronal patterns echoes the compression of neuronal replay events in the hippocampus [9,36].

## Querying the hippocampus with a stable HD signal during SWRs.

During NREM sleep, HD cell activity drifts at angular velocities that are, on average, 5 to 10 times higher than during wakefulness and REM sleep (i.e. active states) yet, this distribution is broad with velocities ranging from stable direction to fast sweeps at angular speed never observed during exploration[25,37]. Interestingly, this echoes animals' natural movements, in which bouts of locomotion associated with fairly stable head direction alternate with pauses during which animals perform head scans [38,39]. Could it be possible that a similar relationship exists between virtual head movement and spontaneous replay of trajectories in the hippocampus during NREM? Furthermore, in contrast to active states during which HD cell population maintains high activity level (i.e. gain), HD cell population often visits "forbidden" states of the attractor during NREM, with the network gain spanning all possible magnitudes from active state levels during UP states to global population silence during DOWN states [24]. As the ADN is a key node in the synchronization of the slow oscillation (see above), one could expect a tight relationship between ADN gain levels and hippocampal activity.

Simultaneous recordings of ADN-HD cell populations and LFP in CA1 show that ADN-HD cells reliably fire immediately before SWRs (Fig. 4A). This indicates that the gain of ADN-HD cells increases prior to SWRs (to note, this gain modulation is independent of DOWN-UP fluctuations as SWRs can occur at any phase of the ADN UP state [18]). The ADN is the only anterior thalamic nucleus to show this homogeneous increase before SWRs [18], pointing to a specific role of this nucleus in coordinating activity across large cortical networks [20]. To decode the HD signal, HD cell population activity was projected onto a 2-dimensional subspace [18,24] (Fig. 3C) in which the radius and the angle from origin correspond to the gain and internal HD, respectively. Two key observations are made. First, as expected from single cell response, the HD cell population trajectories in this projection subspace reach the outer perimeter of the distribution immediately before SWRs, with gain levels corresponding to active states (Fig. 3D). Second, the ADN-HD signal is stabilized (as measured by the decrease in angular velocity), also before SWRs (Fig. 3E). This inverse relationship between drifting speed and gain in the ADN-HD network may not be specific to sleep as it was recently observed in foraging mice [40].

In conclusion, these results raise the intriguing possibility that the content of hippocampal replay is constrained by subcortical inputs, in particular a stable and high-gain HD signal. In other words, the HD system would query the hippocampal network with a particular direction. These observations also solve the conundrum of how fast HD signal drift during NREM can be compatible with veridical replay in the hippocampus. In fact, like during natural movement, replay of trajectories is associated with a stable HD signal. Yet, replayed trajectories after exploration of an open field are not all necessarily linear [41]. The HD signal may thus perhaps constrain the overall direction of the replayed trajectories or perhaps just the starting direction. At any rate, the ADN-PoSub HD network does not contact the hippocampus directly, and the question remains of whether the MEC exhibits similar properties around SWRs.

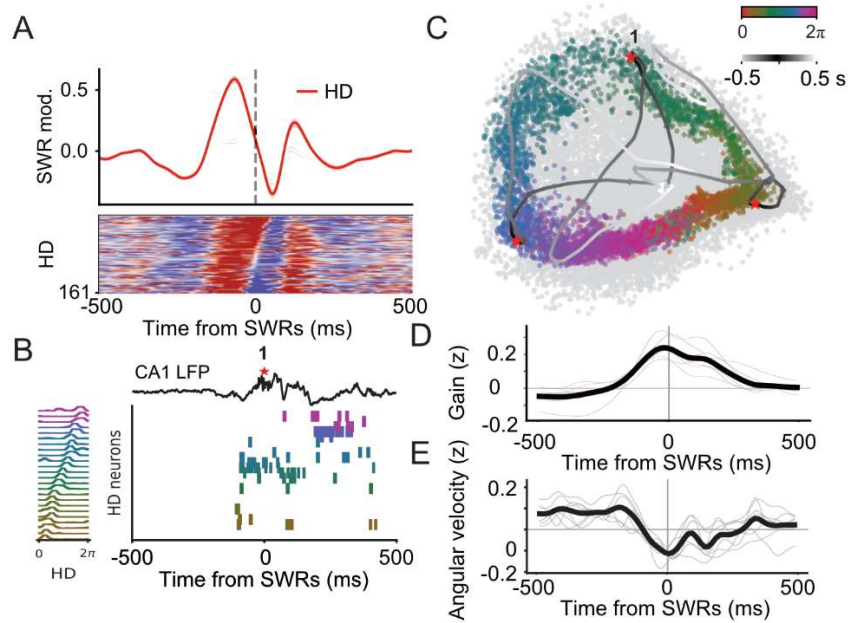

**Figure 3. Increases gain and stabilization of the HD signal immediately before SWRs.**
A) Normalized activation of individual ADN-HD cells relative to SWR times (bottom) and average (top).
B) Example ADN-HD cell population activity around the time of a SWR (indicated with an asterisk in the CA1 LFP trace, top). Cells are sorted relative to their preferred direction (tuning curves in cartesian plots shown in left).
C) ADH-HD cell population trajectories from -500 to 500ms around three example SWRs in 2-dimensional projection space. Projected activity in NREM (gray dots) is shown together with population activity during wakefulness (colored dots). The three asterisks indicate the time of three example SWRs with "1" indicating the one showing in B. Note that population activity reaches gain levels of wakefulness immediately before and during each SWR.
D) Average normalized gain of ADN-HD cells around SWRs (dark line). Each individual session is shown as thin gray line.
E) Same as D for angular velocity of the ADN-HD signal.
Adapted from ref [18]

## Is the content of hippocampal replay dictated by its inputs?

The input-output organization of the hippocampus in the MEC is divided across its layers, with the superficial and deep layers constituting the inputs and output of the hippocampus, respectively [42]. Hippocampal SWRs have a strong impact on its output structures [43], directly activating deep layers of retrohippocampal areas, including the MEC [43]. As grid cells are coordinated during NREM sleep, do they replay trajectories coherently with the hippocampus and is there a difference across layers?

Two studies performing simultaneous recordings in the CA1-MEC networks provided fundamental insights into this question. CA1 place cells and deep layer grid cells are reactivated coherently [44]. Grid cell replay tends to follow place cell replay, as expected from CA1 projections to the deep layers of the MEC [42] and the delayed ripples observed in these layers [43]. In contrast, superficial layer grid cell replay events were largely independent of CA1 replay when considered from a pure spatial perspective (i.e. replay of spatial trajectories in the MEC) [45]. However, despite these observations, it is too premature to declare that hippocampal information content during replay is only dictated by intrahippocampal events. First, a majority of cells in MEC layer 3 – which project to CA1 - show directional tuning [27]. The correspondence between the activity of these MEC HD cells and CA1 replay is still unknown. Furthermore, it is possible that grid cells themselves provide directional input to the hippocampus independently of any specific (i.e. replayed) trajectory. This could be encoded by the drifting direction of grid cells, not necessarily related to the replay of specific neuronal patterns. Hence, considering the dynamics of ADN-HD cells around SWRs and the contribution of HD cell firing for MEC spatial representation [29,46], it is in fact quite likely that a directional signal linked to SWRs activity is present in the MEC, upstream the hippocampus.

## Learning to read out hippocampal code during SWRs.

One crucial question regarding hippocampal replay is how this code is read out by output structures to support memory consolidation. SWRs activate the deep layers of retrohippocampal areas, including the PoSub [43]. One can speculate on the role of CA1 inputs to the PoSub at times of SWRs. The PoSub is ideally situated to compare the direction queried by ADN-HD cells prior to SWRs with the direction of the replayed sequence in CA1. Mismatch between the queried and decoded directions would lead to an adjustment of the network. In other words, this phenomenon would support the training of a decoder reading out the directional tuning of CA1 ensembles, and alignment between the head direction signal (the compass) and hippocampal maps. This idea is supported by an important detail regarding replay of grid cells in the deep layer of the MEC. In the deep layers of the MEC, grid cells are often also modulated by animal's HD [47]. Interestingly, grid-place coherence was correlated with the directional tuning of deep layer grid cells [44], suggesting a tight relationship between place and HD signals during replay. Considering that HD is influenced by specific viewpoints during navigation and that the visual system shows coordinated replay with the hippocampus [48], the query-training sequence may constitute a fundamental process in the formation of spatial memories and scene representation in the brain (Fig. 4).

While it is still unknown if PoSub-HD cells are specifically activated by SWRs, ADN-HD cells are, in contrast, inhibited immediately after SWRs (Fig. 3A). This observation points to a reciprocal interaction between the HD system and the hippocampus, which may have further implications for spatial memory formation. But what are the circuits that support this post-SWRs inhibition in the ADN? In the thalamus, the main source of inhibition arises from the thalamic reticular nucleus (TRN), a layer of inhibitory neurons surrounding the thalamus that is reciprocally connected with many (but not all) thalamic nuclei [49,50]. In the anterior thalamus, ADN is the nucleus showing the strongest reciprocal connections with the TRN [51–53]. As in any thalamocortical networks, deep layer PoSub neurons send feedforward excitatory connections to the ADN [54] as well as collaterals within the thalamic reticular nucleus (TRN) [53]. These connections result in a strong hyperpolarization of ADN neurons when PoSub cells are activated [53]. Hence, SWR-activated PoSub neurons may lead to the inhibition of ADN neurons.

This post-SWRs activation of the TRN may have two meaningful consequences. First, TRN activation triggers spindles [55], oscillations that are believed to be associated with plasticity in thalamocortical networks [56]. In this case, plasticity during SWR-triggered spindles would support training of the place-to-

HD decoder. It is possible that the CA1-PoSub-TRN pathway is also a key circuit supporting the coordination of SWR and spindles observed in the medial cortex [15,19]. Second, the inverse relationship between network gain and drifting speed in the ADN-HD cell population [18,40] suggests that SWRs lead to an indirect rebound in drifting speed and randomly fluctuations in the HD signal after SWRs. These fast drifts are observed in experimental data (Fig. 3B-C) and could potentially select a new direction at random before the next SWR occurs. In support of these two predictions, HD drifts are associated with increased power in the spindle band [24].

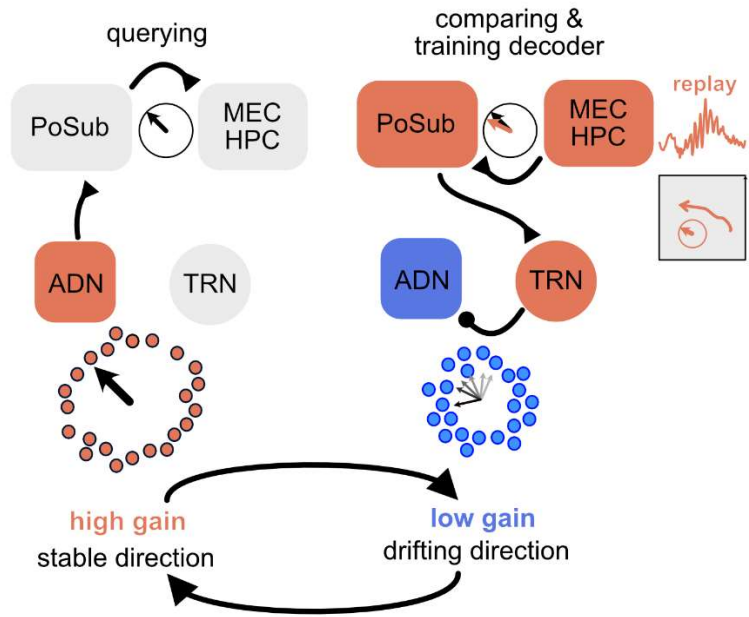

**Figure 4. The querying-training hypothesis.**
During NREM, ADN-HD cell population alternates between states of high and low activity gain, associated with stable and fast drifting directions, respectively (bottom). States of high gain influence downstream structures and may lead to the generation of SWRs in the hippocampus. At times of SWRs, the ADN would query the hippocampus with a specific direction (dark arrow in a circle, top left). During the SWR, a neuronal sequence replays a spatial trajectory (orange trajectory on a schematic square gray box, right), from which a direction can be decoded (orange arrow in a circle). The PoSub can then compare the queried and replayed directions to train a place-to-HD decoder of hippocampal sequences. This process would give meaning to hippocampal sequences and support the formation of long-term spatial memories. In turn, the activation of the PoSub leads to a feedforward inhibition of the ADN (through activation of the TRN), shifting the ADN-HD cell population in a state of low gain and fast drifting direction. Lines connecting structures show excitatory and inhibitory pathways with triangles and circles, respectively.

## Conclusion

In summary, the thalamocortical loop composed of the ADN, PoSub, MEC and the hippocampus would play an essential role in memory formation. The "simple" representation (in David Marr's words) of animal's HD constitutes a key link between the external world and internal state that gives meaning to hippocampal neuronal sequences. Specifically, the PoSub would encode multimodal sensory inputs, especially vision and vestibular information, into a code for HD. In turn, it would decode animal's HD from hippocampal sequences and match it with the estimate provided by sensory evidence. During sleep, the training of this place-to-HD decoder would take place by first querying the hippocampus with a direction, selected by the subcortical HD network, and compare it with the direction decoded from the neuronal sequence that the hippocampus produces during SWRs. Following SWRs, feedforward inhibition of the ADN allows the system to select a new direction.

Of course, similar mechanisms to train the place-to-HD decoder may happen during exploration. However, hippocampal replay during sleep may result from a generative process and not reflect exactly previously experienced journey [41,57]. Hence, spatial memories may be generalized to all possible viewpoints, leading to the formation of a memory for the environment in the cortex.

Whether this mechanism extends to trajectories in non-spatial cognitive space remains open [58,59]. While the HD system is primarily driven by the vestibular system, it is still possible that the ADN-PoSub network disengages from vestibular inputs when the animal navigates abstract cognitive space. It is also possible that SWRs replaying non-spatial events are independent of HD and MEC activity. The lateral entorhinal cortex is another major inputs to the hippocampus [42], and the activity of its neurons as well as its own upstream inputs remain unknown.

## Acknowledgments

This work was supported by a Canadian Research Chair in Systems Neuroscience, CIHR Project Grant 155957, NSERC Discovery Grant RGPIN-2018-04600, and the Canada-Israel Health Research Initiative, jointly funded by the Canadian Institutes of Health Research, the Israel Science Foundation, the International Development Research Centre, Canada and the Azrieli Foundation 108877-001. I thank members of the lab for their comments on an earlier version of this manuscript.

# Bibliography

1. Scoville, W. B. & Milner, B. Loss of recent memory after bilateral hippocampal lesions. *J. Neurol. Neurosurg. Psychiatry* **20**, 11–21 (1957).

2. Burgess, N., Maguire, E. A. & O'Keefe, J. The Human Hippocampus and Spatial and Episodic Memory. *Neuron* **35**, 625–641 (2002).

3. O'Keefe, J. & Dostrovsky, J. The hippocampus as a spatial map. Preliminary evidence from unit activity in the freely-moving rat. *Brain Res.* **34**, 171–175 (1971).

4. Morris, R. G. M., Garrud, P., Rawlins, J. N. P. & O'Keefe, J. Place navigation impaired in rats with hippocampal lesions. *Nature* **297**, 681–683 (1982).

5. Marr, D. Simple memory: a theory for archicortex. *Philos. Trans. R. Soc. Lond. B. Biol. Sci.* **262**, 23–81 (1971).

6. Buzsaki, G. Two-stage model of memory trace formation: a role for 'noisy' brain states. *Neuroscience* **31**, 551–570 (1989).

7. McClelland, J. L., McNaughton, B. L. & O'Reilly, R. C. Why there are complementary learning systems in the hippocampus and neocortex: insights from the successes and failures of connectionist models of learning and memory. *Psychol. Rev.* **102**, 419–457 (1995).

8. Wilson, M. A. & McNaughton, B. L. Reactivation of hippocampal ensemble memories during sleep. *Science* **265**, 676 (1994).

9. Lee, A. K. & Wilson, M. A. Memory of sequential experience in the hippocampus during slow wave sleep. *Neuron* **36**, 1183–1194 (2002).

10. Buzsáki, G. Hippocampal sharp wave-ripple: A cognitive biomarker for episodic memory and planning. *Hippocampus* **25**, 1073–1188 (2015).

11. Girardeau, G., Benchenane, K., Wiener, S. I., Buzsáki, G. & Zugaro, M. B. Selective suppression of hippocampal ripples impairs spatial memory. *Nat. Neurosci.* **12**, 1222–1223 (2009).

12. Gridchyn, I., Schoenenberger, P., O'Neill, J. & Csicsvari, J. Assembly-Specific Disruption of Hippocampal Replay Leads to Selective Memory Deficit. *Neuron* **106**, 291-300.e6 (2020).

13. Ego-Stengel, V. & Wilson, M. A. Spatial selectivity and theta phase precession in CA1 interneurons. *Hippocampus* **17**, 161–174 (2007).

14. Steriade, M., Nunez, A. & Amzica, F. A novel slow (< 1 Hz) oscillation of neocortical neurons in vivo: depolarizing and hyperpolarizing components. *J. Neurosci.* **13**, 3252 (1993).

15. Siapas, A. G. & Wilson, M. A. Coordinated interactions between hippocampal ripples and cortical spindles during slow-wave sleep. *Neuron* **21**, 1123–1128 (1998).

16. Sirota, A., Csicsvari, J., Buhl, D. & Buzsáki, G. Communication between neocortex and hippocampus during sleep in rodents. *Proc. Natl. Acad. Sci. U. S. A.* **100**, 2065 (2003).

17. Isomura, Y. *et al.* Integration and segregation of activity in entorhinal-hippocampal subregions by neocortical slow oscillations. *Neuron* **52**, 871–882 (2006).

18. Viejo, G. & Peyrache, A. Precise coupling of the thalamic head-direction system to hippocampal ripples. *Nat. Commun.* **11**, 2524 (2020).

19. Levenstein, D., Buzsáki, G. & Rinzel, J. NREM sleep in the rodent neocortex and hippocampus reflects excitable dynamics. *Nat. Commun.* **10**, 2478 (2019).

20. Gent, T. C., Bandarabadi, M., Herrera, C. G. & Adamantidis, A. R. Thalamic dual control of sleep and wakefulness. *Nat. Neurosci.* **21**, 974 (2018).

21. Taube, J. S., Muller, R. U. & Ranck, J. B. Head-direction cells recorded from the postsubiculum in freely moving rats. I. Description and quantitative analysis. *J. Neurosci.* **10**, 420–435 (1990).

22. Taube, J. S. The head direction signal: origins and sensory-motor integration. *Annu. Rev. Neurosci.* **30**, 181–207 (2007).

23. McNaughton, B. L., Battaglia, F. P., Jensen, O., Moser, E. I. & Moser, M.-B. Path integration and the neural basis of the 'cognitive map'. *Nat. Rev. Neurosci.* **7**, 663–678 (2006).

24. Chaudhuri, R., Gercek, B., Pandey, B., Peyrache, A. & Fiete, I. The intrinsic attractor manifold and population dynamics of a canonical cognitive circuit across waking and sleep. *Nat. Neurosci.* **22**, 1512–1520 (2019).

25. Peyrache, A., Lacroix, M. M., Petersen, P. C. & Buzsáki, G. Internally organized mechanisms of the head direction sense. *Nat. Neurosci.* **18**, 569–575 (2015).

26. Adamantidis, A. R., Gutierrez Herrera, C. & Gent, T. C. Oscillating circuitries in the sleeping brain. *Nat. Rev. Neurosci.* **20**, 746–762 (2019).

27. Giocomo, L. M. *et al.* Topography of head direction cells in medial entorhinal cortex. *Curr. Biol. CB* **24**, 252–262 (2014).

28. Hafting, T., Fyhn, M., Molden, S., Moser, M.-B. & Moser, E. I. Microstructure of a spatial map in the entorhinal cortex. *Nature* **436**, 801–806 (2005).

29. Winter, S. S., Clark, B. J. & Taube, J. S. Disruption of the head direction cell network impairs the parahippocampal grid cell signal. *Science* **347**, 870–874 (2015).

30. Fuhs, M. C. & Touretzky, D. S. A Spin Glass Model of Path Integration in Rat Medial Entorhinal Cortex. *J. Neurosci.* **26**, 4266–4276 (2006).

31. Burak, Y. & Fiete, I. R. Accurate Path Integration in Continuous Attractor Network Models of Grid Cells. *PLOS Comput. Biol.* **5**, e1000291 (2009).

32. Sorscher, B., Mel, G. C., Ocko, S. A., Giocomo, L. & Ganguli, S. A unified theory for the computational and mechanistic origins of grid cells. 2020.12.29.424583 (2020) doi:10.1101/2020.12.29.424583.

33. Gardner, R. J., Lu, L., Wernle, T., Moser, M.-B. & Moser, E. I. Correlation structure of grid cells is preserved during sleep. *Nat. Neurosci.* **22**, 598 (2019).

34. Gardner, R. J. *et al.* Toroidal topology of population activity in grid cells. *Nature* **602**, 123–128 (2022).

35. Trettel, S. G., Trimper, J. B., Hwaun, E., Fiete, I. R. & Colgin, L. L. Grid cell co-activity patterns during sleep reflect spatial overlap of grid fields during active behaviors. *Nat. Neurosci.* **22**, 609 (2019).

36. Skaggs, W. E. & McNaughton, B. L. Replay of neuronal firing sequences in rat hippocampus during sleep following spatial experience. *Science* **271**, 1870–1873 (1996).

37. Brandon, M. P., Bogaard, A. R., Andrews, C. M. & Hasselmo, M. E. Head direction cells in the postsubiculum do not show replay of prior waking sequences during sleep. *Hippocampus* **22**, 604–618 (2012).

38. Drai, D., Kafkafi, N., Benjamini, Y., Elmer, G. & Golani, I. Rats and mice share common ethologically relevant parameters of exploratory behavior. *Behav. Brain Res.* **125**, 133–140 (2001).

39. Monaco, J. D., Rao, G., Roth, E. D. & Knierim, J. J. Attentive scanning behavior drives one-trial potentiation of hippocampal place fields. *Nat. Neurosci.* **17**, 725–731 (2014).

40. Ajabi, Z., Keinath, A. T., Wei, X.-X. & Brandon, M. P. Population dynamics of the thalamic head direction system during drift and reorientation. *bioRxiv* (2021).

41. Stella, F., Baracskay, P., O'Neill, J. & Csicsvari, J. Hippocampal Reactivation of Random Trajectories Resembling Brownian Diffusion. *Neuron* **102**, 450-461.e7 (2019).

42. van Strien, N. M., Cappaert, N. L. M. & Witter, M. P. The anatomy of memory: an interactive overview of the parahippocampal–hippocampal network. *Nat. Rev. Neurosci.* **10**, 272–282 (2009).

43. Chrobak, J. J. & Buzsáki, G. High-frequency oscillations in the output networks of the hippocampal–entorhinal axis of the freely behaving rat. *J. Neurosci.* **16**, 3056–3066 (1996).

44. Ólafsdóttir, H. F., Carpenter, F. & Barry, C. Coordinated grid and place cell replay during rest. *Nat. Neurosci.* **19**, 792–794 (2016).

45. O'Neill, J., Boccara, C. N., Stella, F., Schoenenberger, P. & Csicsvari, J. Superficial layers of the medial entorhinal cortex replay independently of the hippocampus. *Science* **355**, 184–188 (2017).

46. Tukker, J. J., Tang, Q., Burgalossi, A. & Brecht, M. Head-Directional Tuning and Theta Modulation of Anatomically Identified Neurons in the Presubiculum. *J. Neurosci.* **35**, 15391–15395 (2015).

47. Sargolini, F. *et al.* Conjunctive representation of position, direction, and velocity in entorhinal cortex. *Science* **312**, 758–762 (2006).

48. Ji, D. & Wilson, M. A. Coordinated memory replay in the visual cortex and hippocampus during sleep. *Nat. Neurosci.* **10**, 100–107 (2006).

49. Jones, E. G. *The Thalamus*. (Cambridge Univ Pr, 2007).

50. Sherman, S. M. & Guillery, R. W. The role of the thalamus in the flow of information to the cortex. *Philos. Trans. R. Soc. B Biol. Sci.* **357**, 1695–1708 (2002).

51. Duszkiewicz, A. J. *et al.* Reciprocal representation of encoded features by cortical excitatory and inhibitory neuronal populations. 2022.03.14.484357 (2022) doi:10.1101/2022.03.14.484357.

52. Gonzalo-Ruiz, A. & Lieberman, A. R. Topographic organization of projections from the thalamic reticular nucleus to the anterior thalamic nuclei in the rat. *Brain Res. Bull.* **37**, 17–35 (1995).

53. Vantomme, G. *et al.* A Thalamic Reticular Circuit for Head Direction Cell Tuning and Spatial Navigation. *Cell Rep.* **31**, 107747 (2020).

54. Yoder, R. M. & Taube, J. S. Projections to the anterodorsal thalamus and lateral mammillary nuclei arise from different cell populations within the postsubiculum: Implications for the control of head direction cells. *Hippocampus* **21**, 1062–1073 (2011).

55. Fernandez, L. M. J. & Lüthi, A. Sleep Spindles: Mechanisms and Functions. *Physiol. Rev.* **100**, 805–868 (2019).

56. Peyrache, A. & Seibt, J. A mechanism for learning with sleep spindles. *Philos. Trans. R. Soc. B Biol. Sci.* **375**, 20190230 (2020).

57. Swanson, R. A., Levenstein, D., McClain, K., Tingley, D. & Buzsáki, G. Variable specificity of memory trace reactivation during hippocampal sharp wave ripples. *Curr. Opin. Behav. Sci.* **32**, 126–135 (2020).

58. Aronov, D. & Tank, D. W. Engagement of Neural Circuits Underlying 2D Spatial Navigation in a Rodent Virtual Reality System. *Neuron* **84**, 442–456 (2014).

59. Whittington, J. C. R. *et al.* The Tolman-Eichenbaum Machine: Unifying Space and Relational Memory through Generalization in the Hippocampal Formation. *Cell* **183**, 1249-1263.e23 (2020).